\documentstyle[aps,preprint]{revtex}
\def\be{\begin{equation}}
\def\ee{\end{equation}}
\def\bea{\begin{eqnarray}}
\def\eea{\end{eqnarray}}

\begin{document}
\draft


\preprint{IASSNS-HEP-95-98}
\title {\bf M THEORY AND P-BRANES}

\bigskip

\author{ Jnanadeva Maharana \footnote{e-mail: maharana@sns.ias.edu \\
Permanent Address: Institute of Physics, Bhubaneswar 751005, India} }
\address{School of Natural Sciences, Institute for Advanced Study\\
Olden Lane, Princeton, NJ 08540, USA
}
\maketitle
\begin{center}

November, 1995
\end{center}
\begin{abstract}

Ten dimensional type IIA and IIB theories with p-branes are compactified to
8-dimensions.
It is shown that resulting branes can be classified according to the
representations of $\bf {SL(3,Z) \times SL(2,Z)}$. These p-branes can also be
obtained by compactification of M theory on three torus and various wrappings
of membrane and five brane
of the eleven dimensional theory. It is argued that there is evidence for bound
states of the branes in eight dimensions as is the case in the interpretation
of $\bf {SL(2,Z)}$ family of string solutions obtained by Schwarz.

\end{abstract}
\vspace{2 cm}

\narrowtext

\newpage


Recently, a considerable progress has been made in our understanding of
non-perturbative phenomena in string theory [1-4]. The extended objects, the
`p-branes',
have played a key  role in these developments. They appear as non-perturbative
solutions of the low energy string effective field theory [5,6] and they have
been
instrumental in providing an understanding of duality symmetry conjectures in
string theory. Furthermore, p-branes carrying Ramond-Ramond charges have
important
implications for  investigations of dualities and string dynamics in diverse
dimensions [4,7]. It is now accepted that all the five string theories are
intimately related
and there is only one underlying theory and different theories are
manifestations of various
phases of a unique theory.

The conjectured $\bf {SL(2,Z)}$ symmetry [8,9] of ten dimensional type IIB
superstring
theory has interesting consequences. The strong and weak coupling regimes are
related by the duality similar to $N=4, D=4$ heterotic string. We recall that
in
type IIB theory strings carrying two gauge field charges are related (this
interchange is analogous to T-duality) in contrast
to the four dimensional case where electric and magnetic charge carrying
particle states are
connected. When we consider theory toroidally compactified to less than ten
spacetime dimensions, the $\bf {SL(2,Z)}$ together with T-duality group results
in
U-duality group [8].

 The existence of a family of string solutions of type IIB theory in
10-dimensions has been demonstrated by Schwarz [10] recently. Of special
significance are the
 BPS saturated solutions which form an $\bf {SL(2,Z)}$ multiplet,  and these
are labelled by
a pair of integers $(m,n)$, where m and n are relatively prime.  When the type
IIB
theory is compactified on a circle and the spectrum of the $D=9$ theory is
compared
with eleven dimensional supergravity compactified on a torus, several
interesting
results follow. The $\bf {SL(2,Z)}$ duality of type IIB theory, in $D=10$,
corresponds to the modular group of the torus; moreover, one can interpret type
II
theories as wrapped supermembranes of $D=11$ supergravity. In sequel [11], nine
dimensional type IIB theory on $\bf {R^9 \times S^1}$ with p-branes was
considered
along with M theory on $\bf {R^9 \times T^2}$. The eleven dimensional M theory
admits only a membrane and a five membrane [12]. It was found that the p-brane
tensions of type IIB theory could be related to the tensions of M theory using
simple heuristic arguments. Similar relations were also derived for the
p-branes
appearing in type IIA theory. In view of these developments, it is of interest
to
study p-branes in 10-dimensions, their compactifications to lower dimensions
and
various duality symmetries. We shall show that the study of p-branes in eight
dimensions results in revealing many interesting features which can be
understood
from the perspectives of type II theories as well as that of M theory.

In an interesting paper, Polchinski [13] has shown that p-branes carrying RR
charges can be
described by an exact conformal field theory. Moreover, IIA theory admits even
p-branes
and IIB theory couples to odd p-branes. Witten [14], in a beautiful paper, has
shown that
the $\bf {SL(2,Z)}$ family of string solutions obtained by Schwarz [10] can be
interpreted as the
bound states of strings and D-strings and the existence such of BPS saturated
(m,n) states is equivalent to existence of vacua of $1+1$ dimensional
supersymmetric Yang-Mills theories with a mass gap. Subsequently, several
authors have have addressed the problem of bound states for D-branes [15,16].

 In
ten spacetime dimensions, we recall that for p-branes carrying RR charges, type
IIA theory admits even branes whereas type IIB couples to odd branes. When we
turn to NS-NS sector, both theories admit a string [17] and a five brane [18].
We mention in passing that more detailed discussions of p-branes can be found
in references
[5,6,18]. If we turn our attention to
 the p-branes in 8-dimensions [19], we have not only to take into
account the compactification from $D=10$ to $D=8$, but also consider how the
branes wrap around various geometries. The M theory provides another
perspective
of the p-branes in eight dimensions. Note that the eleven dimensional  theory
admits only membrane and five brane; therefore the p-branes in 8-dimensions
will arise from dimensional reductions and various kinds of wrappings as we
go from $D=11$ to $D=8$. To illustrate the point, let us consider membrane and
4-brane which can arise in ten dimensional IIA theory. When we go to
8-dimensions
the dimensional reduction [20] of membrane will take a membrane to a membrane
whereas
the double dimensional reduction [21,6] of the 4-brane will also result in a
membrane.
Furthermore, the interpretation of these two membranes in $D=11$ M theory is
different: one membrane arises from dimensional reduction of membrane of this
theory and another comes from wrapping of the five brane around three torus.
We shall see that when we consider the p-branes in eight dimensions several
interesting results follow.

As a definite case, let us consider the ten dimensional theory with 4-brane.
The relevant
  the action in ten dimensions that admits 4-brane solution is given by

\bea \tilde  I_{10}(5) = {1\over {2 {\kappa_{10}^2}}}
\int d^{10}x \sqrt {-\tilde g} \bigg\{\tilde R - {1\over 2} (\partial \tilde
\phi)^2 - {1\over 2} {1\over 6!} e^{-{1\over 2}\tilde \phi }\tilde F^2_{6}
\bigg\} \eea

The 4-brane couples to the worldvolume as

\bea \tilde S_5&= &\tilde T_5 \int d^5\xi \bigg\{\sqrt {-\tilde
\gamma}
\tilde \gamma^{ij} \partial _iX^M \partial _jX^N\tilde g_{MN}
e^{{1\over 10}\tilde \phi}
+ {3\over 2}\sqrt {-\tilde \gamma} \nonumber\\&-&{1\over 5!}
 \epsilon^{i_1i_2i_3i_4i_5}\partial_{i_1}X^M\partial_{i_2}X^N
\partial_{i_3}X^P\partial_{i_4}X^Q\partial_{i_5}X^R
\tilde A^{MNPQR} \bigg\} \eea

 The actions $\tilde  I_{10}(5)$ and $\tilde  S_5$ are defined with canonical
metrics; here the fields with tilde refer to objects in ten spacetime
dimensions. Note that
$\tilde F_6$ refers to the 6-form field strength and the corresponding 5-form
gauge potential is denoted by $\tilde A$.
In what follows, we recapitulate the essential steps to obtain brane solution
and its 'dual' solitonic solution, in the context of 4-brane and we refer to
the
review article of Duff et. al [6]
for details. While looking for the 4-brane solution, we split the coordinates
as
$x^M = (x^{\mu}, y^m)$, where $\mu = 0,1,2,3,4$ and $m = 5,6 ....9$
and the metric ansatz is

\bea ds^2 = e^{2\tilde A}\eta_{\mu\nu}dx^{\mu}dx^{\nu} +
e^{2\tilde B}\delta_{mn}dy^m dy^n \eea

 \noindent and the ansatz for the five form gauge potential is $\tilde
A_{\mu_1..\mu_5} = \epsilon_{\mu_1..\mu_5}e^{\tilde C}$. We demand invariance
under Poincare transformations in directions 0,1,2,3 and 4 and $SO(5)$
rotational invariance in y coordinates. Then,
$\tilde \phi$,$\tilde A,\tilde B$ and $\tilde C$ are functions of $y = \sqrt
{\delta_{mn}y^m y^n}$. The
solution we look for is the `electric' 4-brane since we solve the field
equation
for combined action $\tilde I_{10}(5) + \tilde S_5$ and the solution is
interpreted as an elementary brane with $e^{-2\tilde \phi} = ( 1 +{ { k_5}\over
{y^3}})^{1/2}$
and $k_5 = 2 \kappa_{10}^2 {{\tilde T_5}\over {3\Omega_4}}$, $\Omega_4$ being
the volume of four sphere. The `electric' charge is given by

\bea g_5^{(e)} = {1\over {\sqrt 2 \kappa_{10}}} \int _{S^4}
e^{-{1\over 2} \tilde \phi}~ {^*} \tilde F_6 \eea

\noindent where $S^4$ is the four sphere surrounding the 4-brane. Similarly,
the `magnetic'
charge is

\bea g_3^{(m)} = {1\over {\sqrt 2 \kappa_{10}}} \int _{S^6}
\tilde F_6 \eea

\noindent $g_3^{(m)}$ is nonzero when $\tilde I_{10}(5)$ has a solitonic
membrane
solution. The solitonic membrane is obtained by solving the equations of motion
in the absence of source and adopting an ansatz of combined $P_3 \times SO(7)$
invariance just as 4-brane had $P_5 \times SO(5)$ invariance, P refers to
Poincare transformation. The mass per unit volume of the 4-brane is

\bea  M_5 = {1\over {\sqrt 2}} \vert g_5^{(e)} \vert e^{{1\over 4}
\tilde \phi _0} = \sqrt 2 \kappa_{10} \tilde T_5 e^{{1\over 4} \tilde
\phi _0} \eea

\noindent $\tilde \phi _0$ being the asymptotic constant value of dilaton. The
corresponding mass density of the membrane is

\bea \tilde M_3 = {1\over \sqrt 2} \vert g_3^{(m)} \vert e^{-{1\over 4}
\tilde \phi _0} \eea

\noindent where $g_3^{(m)} = {2\pi n}( {\sqrt 2 \kappa _{10} \tilde T_5})^{-1}$
by the Dirac quantisation rule. We mention in passing that the masses and
charges obey the same equality as the supersymmetric case [6] when one chooses
the
ratio of the coefficients of the kinetic energy term and the WZW terms
appearing
in eq.(1) and (2) as adopted here.  The solitonic states are also BPS mass
saturated states. We also note that if $\lambda _5$ is the coupling
constant (now we are in $\sigma$-model metrics) associated with 4-brane and
$\lambda _3$ is the one for membrane then, one can check that the relation
$(\lambda_ 5)^5 =
 (\lambda _3)^{-3}$ holds.

Let us proceed to envisage the scenario in 8-dimensions when we adopt double
dimensional
reduction to obtain a membrane from the 4-brane; however, the membrane, when it
is dimensionally reduced will be a membrane in eight dimensions. It is
convenient to use a prescription where the determinant of the five-world-volume
metric is equal to the determinant of the three-world-volume metric which we
get after the reduction. The 8-dimensional action takes the following form

\bea I_8(3) = {1\over {2 {\kappa_8 ^2}}} \int d^8x \sqrt {-g} \bigg\{R -
{1\over 2}(\partial \phi)^2 - {1\over 2}{1\over 4!} e^{- \phi} F_4^2 \bigg\}
\eea

\noindent The membrane source term becomes

\bea S_3&= & T_3 \int d^3\xi \bigg\{\sqrt {-\gamma} \gamma^{ij} \partial _iX^M
\partial _jX^N g_{MN} e^{{1\over 3} \phi} + {1\over 2}\sqrt {-\gamma}
\nonumber\\&-& {1\over 3!} \epsilon ^{ijk} \partial _iX^M \partial _jX^N
\partial _kX^P A_{MNP} \bigg\} \eea

Now the spacetime indices take values $M,N=0,1,2,5,6,7,8,9$ (we compactified
$x^3$ and $x^4$) and the world volume indices
run over 0,1 and 2. In eq.(8), $F_4$ refers to the 4-form field strength
associated with the 3-form potential appearing in (9) which arises as
dimensional reduction of 5-form potential, $\tilde A$, in ten dimension. The
membrane tension $T_3$ appearing in the above equation is
proportional to $\tilde T_5$ with a two-volume factor as we come down from five
dimensional world volume to three dimensional one. The constant $\kappa_8^2 =
(2\pi R)^2 \kappa_{10}^2$ where $R$ is the radius of the circles along
directions $x^3$ and $x^4$. The `electric' charge (we still have the source
term) is given by $g_3^{(e)}=(\sqrt 2 \kappa_8)^{-1} \int _{S^4} e^{-\phi}
{^*}~ F_4$ and the magnetic charge is
$g_3^{(m)}=(\sqrt 2 \kappa_8)^{-1} \int _{S^4} F_4$ and they satisfy the
quantization condition $g_3^{(e)} g_3^{(m)}=2\pi n$, where n is an integer. The
solution to the
field equation can obtained in a straight forward manner by demanding
invariance
under   $P_3\times SO(5)$ transformations  and choosing appropriate ansatz for
the break up of the metric (now $x^{\mu}$ take three values and $y^m$ go over
five
values) and taking the gauge potential to be 1 to three index $\epsilon$ tensor
times a function of y. It is evident that we shall have electrically and
magnetically charged membranes.

Now let us turn our attention to the study of p-branes in 8-dimensions and look
for their origin in type IIB theory in ten dimension, a path taken by Schwarz
[10]
while considering strings in IIB theory in ten and nine dimensions. The type
IIB theory, in RR sector has 3-form and 1 5-form field strengths thus
admitting a string and a 3-brane; in addition a five brane ( we shall refrain
from considering higher p-branes, $p>7$, here). The NS-NS sector has a string
and a 5-brane
in D=10. We note that with each p-brane we can associate a $p+2$ form field
strength. Again, to be specific, let us look at the 4-branes in $D=8$. When we
consider RR sector, the 7-form field strength $\tilde H^{(7)}$ will give rise
to two 6-form field strengths $H^{(6)}_{\alpha}$, $\alpha =3,4$ being compact
directions. From the NS-NS sector, we have four 6-form field strengths: two
coming from the 7-form in NS-NS sector and the other two come from the dual
of two of the 2-form field strengths (which come from dimensional reduction of
3-form field strength in ten dimensions). Thus there are altogether six 6-form
field strengths and we conclude that {\it in eight spacetime dimensions there
are six 4-branes
}. There is another way to cross check our accountings. We know, when the
type IIB theory is toroidally compactified to 8-dimensions, there are six gauge
bosons: two pairs come from the dimensional reduction of the ten dimensional
metric and antisymmetric tensor
fields (these are from NS-NS sector) and two more from the antisymmetric tensor
of
the RR sector. Notice that the existence of the six gauge bosons implies that
there are six 0-branes. Since, in $D=8$, the dual of a 0-brane is a
four brane, we should have six 4-branes.  Since IIB theory has two
five branes (each from NS-NS and RR sector), in ten dimensions, the
four branes will arise when the five branes wrap around the two torus
to give rise to the 4-branes we have been discussing.  How do we
understand these 4-branes from eleven dimensional M theory?  There is
one five brane in D=11 and we come down to a D=8 compactifying on three torus.
The 4 will give
three of the 4-branes and the rest will come from the KK modes. Now to complete
the discussion of the 4-branes, let us 2 their origin from type IIA view
point since IIA and IIB theories are equivalent in 8-dimensions. In ten
dimensions, the graviton and antisymmetric tensor field originate
from NS-NS sector and the lone gauge field and 3-form potential arise from the
RR sector. When we count the number of 6-form field strengths in $D=8$, they
add up to
six. We can follow same line of arguments for counting of other branes
in 8-dimensions.

There are interesting consequences of these results. It is well known that when
IIB theory is compactified to 8-dimensions, the resulting fields can be grouped
into $\bf {SL(3,) \times SL(2,)}$ representations. The gauge six gauge fields,
alluded to above, belong
to $\bf {(3,2)}$ representations and consequently, the 0-branes and their duals
the
4-branes also belong to this representations. We can adopt and generalize the
arguments of Schwarz [10], and propose that these 4-branes/0-branes will carry
$\bf {SL(3,Z)
\times SL(2,Z)}$ charges, and those objects carrying charges with relatively
prime integer
 will be stable. Such 4-branes/0-branes can be interpreted as bound states of
other branes. We know that as we come to $D=8$, the group is product of $\bf
{SL(2,Z)}$, which was in ten dimensions and $O(2,2;\bf Z)$ which arises as a
result of
dimensional reduction. The $O(2,2;\bf Z)$ has $\bf {SL(2,Z)\times SL(2,Z)}$ as
its subgroup. One of these $\bf {SL(2,Z)}$, the one which parametrizes $
B_{34}+i
\sqrt {det G_{\alpha \beta}}, \alpha, \beta=3,4$ combines with the $\bf
{SL(2,Z)}$
coming from $D=10$ and the $\bf {SL(3,Z)}$ is a subgroup of the product of
these two $\bf {SL(2,Z)}$ groups. Here $B_{34}$ and $G_{\alpha \beta}$ refer to
the internal components of antisymmetric tensor
and the metric as we come from ten to eight dimensions.

When we turn out attentions the membranes in 8-dimensions, we find that there
is only a pair  of them. This can
be seen by counting 4-form fields, after dimensional reduction, either in
type IIA, or in IIB or in M theories. It is easy to see that each of the theory
contains only two such field
strengths. Therefore, we conclude that the 4-form field strength $F_4$ and its
dual ${^*}F_4$ belong to
 $(1,2)$ representation of $\bf {SL(3,Z) \times SL(2,Z)}$.
 Indeed, the membranes will be characterized by
a pair of integers $(m,n)$ and the dyonic solutions of Izquierdo et. al [18]
now finds a natural interpretation
in this perspective. Now we can invoke the arguments of Schwarz, [10] for $\bf
{SL(2,Z)}$ family of strings, and claim that if a membrane carries charges
$(m,n)$, m and n relatively prime, they will be stable. Therefore, bound states
of membrane should
exist as stable membranes.

 It follows from results of ref.11  that the p-brane tensions are related to
the membrane and fivebrane tensions of M theory.
It is well known that the  IIA theory has a simple interpretation as the M
theory
on $R^{10} \times S^1$.
 If $g^{(M)}$ is the eleven dimensional metric,
and $ L=2\pi R$ is the circumference in that metric, then the string metric,
$g^{(A)}$, of type IIA theory is $g^{(A)}=e^{2\phi_A /3} g^{(M)}$ and the
dilaton of the IIA theory is identified as $\phi _A$ and the coupling constant
is the
vacuum expectation value of $e^{\phi _A}$. The set of relations derived by
Schwarz [11]
are ($T_1^{(A)}$, $T_2^{(A)}$ and $T_4^{(A)}$ denote IIA theory tensions for
string, membrane and 4-brane in what follows and similar parameters with
superscript M refer to the M-theory counter parts):

\bea T_2^{(A)} = g_A^{-1} T_2^{(M)},~~~~~
 T_4^{(A)} = g_A^{-{5\over 3}} LT_5^{(M)} \eea

\noindent for the even p-branes coming from the RR sector and

\bea T_1^{(A)} = g_A^{-{2\over 3}} LT_2^{(M)}, ~~~~~
 T_5^{(A)} = g_A^{-2} T_5^{(M)} \eea

\noindent for string and 5-branes in the NS-NS sector. Therefore,
when we come down to $D=8$, the tensions of four branes and two branes can
be expressed in terms of M theory tensions and the volume factors. It is quite
interesting to see all the intimate connections not only between IIA and IIB
theories,
but also with M theory in the 8-dimensional world.

It is evident that we can consider theories compactified to lower dimensions
and systematically study p-branes in those theories starting from type II
theories
or M theory. Then, dualities and classification of 0-branes, strings and
membranes
can be studied by adopting this procedure. Recently, membrane solutions have
been
obtained for type IIA and heterotic strings in 6-dimensions. It is natural to
expect
that there will be many more solutions than the types of solutions obtained by
Johnson et. al [22].

To summarize, we have studied p-branes in 8-dimensional theory from
compactification
of ten dimensional type II theories. They can be viewed from the M theory
perspective, where the eleven dimensional theory is compactified to
8-dimensions.
The appearance of branes, classified according to $\bf {SL(3,Z)\times
SL(2,Z)}$,
tells us that we expect to have stable 0-branes as well as 4-branes when their
charge assignments satisfy suitable constraints and from these considerations,
we can conjecture that there is evidence for stable bound states of such
branes.
Our results can also be viewed as a way to verify the predictions of U-duality
[8]. Furthermore, we have shown  intimate connections between the
type II theories and the M theory in these studies of p-branes and we
demonstrated
simple and elegant descriptions of the eight dimensional p-branes from the view
point of M theory.

\noindent{\bf Acknowledgement}

\noindent  I would like to thank E. Witten for valuable discussions. I am
thankful
to John Schwarz for sharing his insights of M theory and for his constant
encouragements throughout this work and to K.Intriligator for carefully reading
the manuscript. The gracious hospitality of E. Witten and the
Institute for Advanced Study is gratefully acknowledged. This work is supported
by NSF Grant PHY 92-45317.


\begin{thebibliography}{99}

\bibitem{1} E. Witten, ``Some Comments on String Theory Dynamics''; Proc.
String '95, USC, March 1995, hep-th/9507121.
\bibitem{2} A. Sen, Int. J. Mod. Phys. A9(1994)3707, hep-th/9402002.
\bibitem{3} J. H. Schwarz, Lett. Math. Phys. 34(1995)309, hep-th/9411178.


\bibitem{4} E. Witten, Nucl. Phys. B443(1995)85, hep-th/9503124.
\bibitem{5} C. G. Callan, J. A. Harvey and A. Strominger, Proc. of 1991 Trieste
Spring School on String Theory and Quantum Gravity; G. T. Horowitz, Proc. of
1992 Trieste Spring School on String Theory and Quantum Gravity,
hep-th/9210119; P. K. Townsend, ``p-brane Democracy'', Proc. PASCOS/ Hopkins
Workshop, March 1995, hep-th/9507048.

\bibitem{6} M. J. Duff, R. R. Khuri and J. X. Lu, Phys. Rep. C259(1995)213,
hep-th/9412184.

\bibitem{7}  P. K. Townsend, Phys. Lett. 350B(1995)184 hep-th/951068.

\bibitem{8} C. Hull and P. K. Townsend, Nucl. Phys. B438(1995)109,
hep-th/9410167.

\bibitem{9} M. B. Green and J. H. Schwarz, unpublished.
\bibitem{10} J. H. Schwarz, Phys. Lett. 360B(1995)13,  hep-th/9508143;
J. H. Schwarz,  ``Superstring Dualities'' CALT-68-2019, hep-th/9509148.
\bibitem{11} J. H. Schwarz, ``The Power of M Theory'',  RU-95-68, CALT-68-2025,
hep-th/9510086.
\bibitem{12} M. J. Duff and K. S. Steele, Phys. Lett. 253B(1991)113; R. Guven,
Phys. Lett. 276B(1992)49.
\bibitem{13} J. Polchinski, ``Dirichlet Branes and Ramond-Ramond Charges'',
hep-th/9510017.
\bibitem{14} E. Witten, ``Bound Sates of Strings and p-branes''
IASSNS-HEP-95-83, hep-th/9510135.
\bibitem{15} M. Bershadsky, V. Sadov and C. Vafa, ``D-Strings and D-Manifolds,
hep-th 9510225; M. Li, ``Boundary States of D-branes and Dy-Strings'',
hep-th/9510161.
\bibitem{16} A. Sen, A Note on Marginally Stable Bound States in Type II String
Theory, hep-th/9510225 and ``U-duality and Intersecting D-branes'',
hep-th/9511026.
\bibitem{17} A. Dhabolkar, G. W. Gibbons, J. A. Harvey and F. Ruiz Ruiz, Nucl.
Phys. B340(1990)33; A. Dhabolkar and J. A. Harvey, Phys. Rev. Lett.
63(1989)478.
\bibitem{18} A. Strominger, Nucl. Phys.B343(1990)167; C. G. Callan, J. A.
Harvey and  A. Strominger, Nucl. Phys. B359(1991)611;
 G. T. Horowitz and A. Strominger, Nucl. Phys.B360(1991)197; M. J. Duff and J.
X. Lu, Phys. Lett. 253B(1991)409.
\bibitem{19} p-branes in 8-dimensions have been discussed in M. J. Duff and J.
X. Lu, Nucl. Phys. B416(1994)301; E. Bergshoeff, H. J. Boonstra and T. Ortin,
``S-duality in type II String Theory'', hep-th/9508091; P. K. Townsend, Phys.
Lett. 354B(1995)247; J. M. Izquierdo, N. D. Lambert, G. Papadopoulos and P. K.
Townsend, ``Dyonic Membranes'', hep-th/9508177.
\bibitem{20} J. Maharana and J. H. Schwarz, Nucl. Phys. B390(1993)4; S. Hassan
and A. Sen, Nucl. Phys. B375(1993)103. For a recent review see A. Giveon, M.
Porrati and E. Ravinovici, Phys. Rep. C244(1994)77.
\bibitem{21} M. J. Duff, P. S. Howe, T. Inami and K. S. Stelle, Phys. Lett.
191B(1987)70.
\bibitem{22} C. V. Johnson, N. Kaloper, R. R. Khuri and R. C. Myer, ``Is String
Theory a Theory of Strings?'', hep-th/9509070.

\end{thebibliography}
\end{document}